\documentstyle[prl,aps,epsf,psfig,multicol]{revtex}
\begin{document}
\draft
\title{
{
Non-local effects in the fermion Dynamical mean field framework.\\
Application to the $2D$ Falicov-Kimball model.}}

\author{Mukul S. Laad$^1$ and Mathias van den Bossche$^2$}
\address{$^1$ Institut f\"ur Theoretische Physik
Universit\"at zu K\"oln, D-50937 K\"oln, Germany\\
$^2$  Labortoire de Physique quantique, Universit\'e Paul Sabatier, 
31062 Toulouse cedex, France} 
\maketitle

\begin{abstract}
We propose a new, controlled approximation scheme that explicitly includes
the effects of non-local correlations on the $D=\infty$ 
solution.  In contrast to usual $D=\infty$, the selfenergy is 
selfconsistently coupled to two-particle correlation functions.
The formalism is general, and is applied to the two-dimensional Falicov-Kimball
model.
 Our approach possesses all the strengths of the large-$D$ 
solution, and allows one to undertake a systematic study of the effects of 
inclusion of ${\bf k}$-dependent effects on the $D=\infty$ picture.
Results for the density of states $\rho(\omega)$, and the single particle 
spectral density for the $2D$ Falicov-Kimball model always yield positive
 definite $\rho(\omega)$, and the spectral function shows striking new features 
inaccessible in $D=\infty$. Our results are in good agreement with the 
exact results known on the $2D$ Falikov-Kimball model.
\end{abstract}

\begin{multicols}{2}
\narrowtext

The effects of strong correlations on the properties  of 
low-dimensional lattice fermion systems is 
still an open problem, inspite of many efforts spanning over 
thirty years.  Recently, the development of a {\it dynamical} 
mean-field theory (DMFT), exact in $D=\infty$ has led to a major advancement 
in our understanding 
of the physics in the local limit\cite{hubbard,lisa}. DMFT is a nonperturbative
scheme.  It has provided a
detailed picture of the Mott transition in this limit, and has been 
fruitfully applied to a large class of models.  Moreover, it has proved 
to be a rather good approximation to actual three-dimensional  
transition-metal and rare-earth compounds\cite{lisa}.  
Inspite of its successes, DMFT has its shortcomings; the single particle 
selfenergy is ${\bf k}$-independent, and so is not coupled to 
${\bf k}$-dependent collective 
excitations. This is clearly serious, e.g. near a magnetic 
phase transition where it cannot show any precursor effects.
A consequence of the above is that 
the single particle spectral function 
cannot access such effects, and so the approach cannot be employed to describe 
e.g. angle-resolved photoemission experiments, or to study the changes in Fermi 
surface topologies driven by interactions in finite-$D$.  The above makes it 
imperative to develop such controlled extensions of DMFT as are able to
rectify its unphysical aspects, while preserving its strengths.

The DMFA
 maps the lattice
system onto a selfconsistently 
embedded ``impurity in a bath'' problem that describes the effects of the 
coupling to
 other 
lattice sites. As in a Weiss mean 
field theory, a self consistency condition is obtained by 
requiring the averaged Green's function of the bath $G_{0}$ to coincide
with the local $G$ at this impurity site.
DMFA becomes exact in $D=\infty$ because one can show that 
the spatial correlations fall off at least as $1/D^{|i-j|/2}$ with      
distance $|i-j|$ \cite{vollhardt}. 
 This ``Weiss field'' still has a nontrivial dynamics
described by an effective local action $S_{DMF}$ obtained by integrating 
out all sites excluding the impurity site. Given $S_{DMF}$, one can 
compute the one particle Green's function on this site, $G_{imp}(\omega)$, 
which is a functional of $G_0(\omega).$ 
The selfconsistency condition that links $G_{imp}$ back to $G_{0}$
is solved iteratively. The main drawback of this method 
is that $\Sigma$ is 
${\bf k}$-independent. 

\noindent

 In this letter, we present an extension of the DMFT that allows
one to treat effects of non-local spatial correlations. To do so, we 
exploit the freedom involved in choosing the input bath propagator.
We account for non-local correlations via the self consistent inclusion 
of two particle correlation function in the bath Green's function (GF) 
$G_0(\omega)$. This is achieved by using the Spectral Density Approximation 
(SDA)\cite{roth}. Thus, our aim is to show that a proper combination of 
two methods 
--DMFT and SDA-- is eminently suited to describe $1/D$ effects in lattice 
correlated fermionic systems. As far as application is concerned, we deal 
with the FKM as it was the 
first model to be solved exactly in $D=\infty$. Attempts at a $1/D$ 
expansion for the FKM have also been made. Finally, there exist 
some exact results on the $2D$ FKM which are of great help to evaluate
the quality of our approximation, making it a first choice 
for testing methods in the context of extensions of the DMFT.

We now present this new method. We begin with the formalism for the 
generic Hubbard model at zero-temperature.  We then go to the FK to solve 
the full set of selfconsistent equations.
We will focus on half-filling, the extension away from $n=1$ is straightforward.  The  
 Hamiltonian is
$$
H = -\sum_{<ij>,\sigma}t_\sigma(c_{i\sigma}^{\dag}c_{j\sigma}+h.c) 
+ U\sum_{i}n_{ i\uparrow}n_{i\downarrow} - \mu\sum_{i\sigma}n_{i\sigma}
$$
\noindent on a $D-$dimensional lattice. The Hubbard model is obtained 
when $t_\uparrow = t_\downarrow = t$, and the FKM is 
given by $t_\uparrow = t$ and  $t_\downarrow=0$. In the case of half filling
$\mu=U/2$ by particle-hole symmetry.  These models have been extensively 
studied using the DMFT. 

 We get a bath propagator which includes $1/D$ effects using the SDA, 
first pioneered by Roth \cite{roth} to describe magnetism in narrow
bands. Its  basic idea is to compute exactly the first few
moments of the spectral density to reconstruct the spectral function 
$ A({\bf k},\omega)$.  With H as above, it is easy to 
compute the first four moments. They are given by the expectation value
of the following $n$-fold commutator
\begin{equation}
M_{{\bf k}\sigma}^{(n)}=<[...[[c^\dagger_{{\bf k},\sigma},H],H],...H]>
\end{equation}
Using usual scaling arguments \cite{vollhardt}, one sees that the higher 
$n$, the higher will be the powers of $1/D$ involved in the $n$-th moment. 
In the case of a generic Hubbard model, it turns out that the 
$M_{{\bf k}\sigma}^{(3)}$ contains all the $o(1/D)$ contributions. It 
can be writen $M_{{\bf k}\sigma}^{(3)}=o(\frac{1}{D^0})+U^2n_{\downarrow}(1-n_{\downarrow})B_{{\bf k}\sigma}$ with

\begin{eqnarray}
n_{\downarrow}(1-&&n_{\downarrow})B_{{\bf k}\sigma}
=\frac{1}{N} \sum_{<ij>} 
\{t_{-\sigma} <c_{i-\sigma}^{\dag}c_{j-\sigma}(1-2n_{i\sigma})> \nonumber \\
&& - t_\sigma [<n_{i-\sigma}n_{j-\sigma}>-n_{-\sigma}^{2}-<c_{i\sigma}^{\dag }
c_{i-\sigma}^{\dag}c_{j-\sigma}c_{j\sigma}>\nonumber \\
&& -<c_{i\sigma}^{\dag}c_{j-\sigma}^{\dag}c_{j\sigma}c_{i-\sigma}>]
e^{-i{\bf k}\cdot({\bf R_{i} }-{\bf R_{j}})}\}
\end{eqnarray}

The SDA then permits one to write an explicit closed form expression for $A_\uparrow({\bf k},\omega)=\delta[\omega-\epsilon_{\bf k}+\mu-\Sigma_{0,\uparrow}({\bf k},\omega)]$, where $\Sigma_{0,\uparrow}({\bf k},\omega)$ is given -- for the generic Hubbard model -- by
$$
\Sigma_{0,\uparrow}({\bf k},\omega)=Un_{\downarrow}+U^{2}n_{\downarrow}
(1-n_{\downarrow})[\omega+\mu-U(1-n_{\downarrow})-B_{{\bf k}\downarrow}]^{-1}
$$
This is the self energy in the SDA. It is interesting to notice that 
all the order $1/D$ effects enter only through $B_{{\bf k}\sigma}$,
whereby the self energy  acquires a non-trivial ${\bf k}$-dependence.
At this stage, this quantity is undetermined, and will have to be obtained 
self consistently. The SDA one particle Green's function can then be computed 
as $ G_{0\uparrow}({\bf k},\omega) = [\omega-\epsilon_{\bf k}-
\Sigma_{0,\uparrow}({ \bf k},\omega)]^{-1}$. 

We now consider the FKM, ($t_\downarrow=0$). In this case, the model has a 
local $U(1)$ symmetry,  $[n_{i\downarrow},H]=0$ $\forall i$. It thus follows 
from Elitzur's theorem that only the second correlator in 
$B_{{\bf k}\sigma}$ is non zero. This leads to 
\begin{eqnarray} 
n_{\downarrow}(1-n_{\downarrow})B_{{\bf k}\downarrow}&&=
-\frac{t}{N}\sum_{<ij>}e^{-i{\bf k} \cdot({\bf R_{i}}-{\bf R_{j}})}
[<n_{i\downarrow}n_{j\downarrow}>-n_{\downarrow}^{2}]\nonumber\\ 
&&=-t\chi_{\downarrow}({\bf k}) \label{first} 
\end{eqnarray} 
\noindent where $\chi_{\downarrow}({\bf k})$ is the order $1/D$ {\it static} 
structure factor of the down spin electrons.

\noindent 
 
 It is interesting to notice that the {\it only} order $1/D$ contribution 
to the single particle self energy is in fact $\chi_\downarrow({\bf q})$. 
The above equations are exact in the band as well as in the atomic limit, as 
can be easily checked. This has important 
consequences, for it points to a way for further improvement of the formalism.  
It allows one to formulate a DMFT for the local part, more or less along same
lines as in $D=\infty$ \cite{lisa}. However, in contrast to the regular DMFT 
where $\Sigma_{0\sigma}=0$, $G_0$ contains  information about non-trivial 
${\bf k}$-dependent features. Therefore, using this SDA $G_{0\sigma}$ as the 
input one particle bath Green's function in a new DMFT scheme allows one to 
introduce  in a {\it simple} way the non-local spatial fluctuations at order 
$1/D$ in the FKM.

 We now come to the computation of the dynamical correlations. This 
 results in a dynamically corrected self energy. To do this
beyond the single-site level is a problem that has not been attempted, and 
here we use the standard DMFT as an approximation to the 2D case.  
 In the case of the half filled FKM, the method drawn from the lines
sketched above can be developed the following way.
 
  As is known \cite{BM,laad}, the 
equation-of-motion method (EOMM) gives the exact solution to the FKM in 
$D=\infty$. We will use the EOMM to get the full (local) impurity
Green's function which we denote by $G_{imp,\uparrow}(\omega)$. 

The closed set of equations for the on-site Green's function
with a bath having $\Sigma_{0,\uparrow}({\bf k},\omega)$ as a selfenergy is 
given in ref. \cite{laad}, replacing $\epsilon_{\bf k}$ by $\epsilon_{\bf k}+
\Sigma_{0,\uparrow}({\bf k},\omega)$.

The local $G_{\uparrow}$ is computed to have the same form as that found in 
\cite{lisa}, and the local self-energy is computed from Dyson's equation,
$\Sigma_{imp,\uparrow}(\omega)=G_{0,imp,\uparrow}^{-1}(\omega)-G_{imp,\uparrow}^{-1}(\omega)$ where
$ G_{0,imp,\uparrow}(\omega)=[\omega + \mu - \Delta(\omega)]^{-1} $, with 
\begin{equation}
\Delta(\omega)=\sum_{\bf k} \frac{t_{\bf k}^2}{\omega+\mu- \epsilon_{\bf k}-\Sigma_{0,\uparrow}({\bf k},\omega)}
\end{equation}
One then gets 
\begin{equation}
\Sigma_{imp,\uparrow}(\omega) = Un_\downarrow + \frac{U^2n_\downarrow(1-n_\downarrow)}{\omega + \mu -U(1-n_\downarrow) - \Delta(\omega)}.
\end{equation}

\noindent 

This local dynamical self energy is used to correct the local part of 
the bath self energy, to introduce dynamical effects in the ${\bf k}$-dependent
bath GF. This is done by 
\begin{equation}
\Sigma_{\uparrow}({\bf k},\omega)=\Sigma_{imp,\uparrow}(\omega)+\Sigma_{0,\uparrow}({\bf k},\omega)- \sum_{\bf k}\Sigma_{0,\uparrow}({\bf k},\omega)
\end{equation}
and provides us with a dynamically corrected non-local Green's function
$G_\uparrow({\bf k},\omega)=[\omega+\mu- \epsilon_{\bf k}
-\Sigma_\uparrow({\bf k},\omega)]^{-1}$ which is exact in both the atomic 
and the band limits. The density of states is then obtained by taking the 
imaginary part of the ${\bf k}$-summed $G({\bf k},\omega)$.

 To complete the procedure, one should selfconsistently compute 
the order $1/D$ correlator which enters the bath Green's function
as anounced earlier. In the case of the FKM, this means one has to compute the
down spin static susceptibility.
To begin with, we compute  the vertex function exactly in $D=\infty$. The full 
$\chi_{\downarrow}({\bf q})$ is then obtained from the Bethe-Salpeter 
equation in the framework of the EOMM \cite{BM}.  The result is
$\chi_{\downarrow}({\bf q})=n_{\downarrow}(1-n_{\downarrow})/D({\bf q})$,
where 
\begin{equation}
D({\bf q})=1-\sum_{\nu}\frac{dn_{\downarrow}}{dG_{\nu}^{-1}}
\left .
\frac{\partial \Sigma_\nu}{\partial n_\downarrow}
\right |_{G_\nu}
\frac{G_{\nu}^{2}+ \chi_{\uparrow}^{0}({\bf q})}
{G_{\nu}^{2}[\chi^0_{\uparrow}({\bf q})\frac{d\Sigma_{\nu}}{dG_{\nu}}+1]}
\label{SC1}
\end{equation}
and $ \chi^0_{\uparrow} ({\bf q})=(-1/N)\sum_{\bf k}G_{\uparrow}({\bf k}+{\bf q})G_{\uparrow}({\bf k})$,
where the $\nu$-sum is on the Matsubara indices (we take the zero temperature limit). 
In the above, 
the vertex corrections in the equation for the charge susceptibility of 
the down-spins electrons are approximated by their $D=\infty$ values.
To compute the vertex function exactly to  order $1/D$ would be a 
formidable task, which has not been successfully attempted even for the electron gas.
\begin{figure}
\centerline{\psfig{file=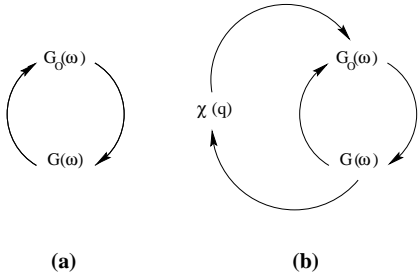,height=4.5cm,angle=0}}
\caption{
(a) The selfconsistency loop in usual DMFT. (b)~Our method:
${\bf k}$-dependent features are incorporated via an external input
$\chi_\downarrow$ to the bath Green's function $G_0$, which is
computed self consistently as shown.}
\end{figure}  
To sum up,  the equations (\ref{first})-(\ref{SC1}) above form a {\it complete} 
set of selfconsistent 
equations for the FKM. These include explicit $1/D$ effects through the 
${\bf k}$-dependence of $\Sigma_{0,\uparrow}({ \bf k},\omega)$.  To
solve these equations, we start with  an arbitrary 
$\chi_{\downarrow}({\bf q})$ as an input in our initial $G_{0\uparrow}$.
This is used to compute $\Sigma_{0,\uparrow}({\bf k},\omega)$, and the full 
$G_{imp,\uparrow}(\omega)$ from the large-$D$ solution along the lines of
\cite{laad}.
At this step, we have 
a DMF approximation to the FKM that is equivalent to a selfconsistent 
loop, with an 
external input quantity, the structure factor. The second step of our
method is to compute $\chi_{\downarrow}({\bf q})$. For this
purpose, we use the DMF approximation to the structure factor equation (\ref{SC1}), plugging the obtained 
$G_{\uparrow}({\bf k},\omega)$ along with the full {\it local} 
vertex function in the Bethe-Salpeter equation to compute 
$\chi_{\downarrow}({\bf q})$. This is fed back into the equation for 
$\Sigma_{0,\uparrow}$ and the whole process is iterated to convergence. 
The difference between the regular self consistent DMFT and our 
method is illustrated in fig.1.

We implemented numerically this new method to study the 
Falicov-Kimball model on a $2D$ square lattice at $T=0$. The up spin electron
dispersion relation is thus 
$\epsilon_{\bf k}=-2t (\cos{k_x} + \cos{k_y})$,
and the corresponding density of states has a van Hove singularity 
at $\omega=0$. The ${\bf k}$ sums are performed on a $L \times L$ 
mesh, while the $\omega$-integrals are computed using standard 
integration procedures, using a small broadening $\eta$ to represent
$\delta$ functions. Convergence is obtained when $\chi^{2}=
\sum_{\omega}|G_{new}(\omega)-G_{old}(\omega)|^{2} < \varepsilon$
where $\varepsilon$ is a small threshold. 
At each step of the iterative process, a check on the accuracy of the 
computation is provided by the Luttinger sum-rule 
$\sum_{\bf k}\int_{-\infty}^{\mu}{\mbox d}\omega A_{\sigma}({\bf k},\omega)=
n_{\sigma}$, which must be satisfied to high accuracy.

We now describe the results of our computation. Working with 
a lattice size $L=64$, and a threshold $\varepsilon=10^{-9}$, we 
observe a very fast convergence of the double selfconsistent procedure,
that needs less than 10 steps to reach the convergence threshold 
for $G_{\uparrow}(\omega)$. The Luttinger sum rule is satisfied 
with an error of less than $10^{-3}$.
\begin{figure}
\centerline{\psfig{file=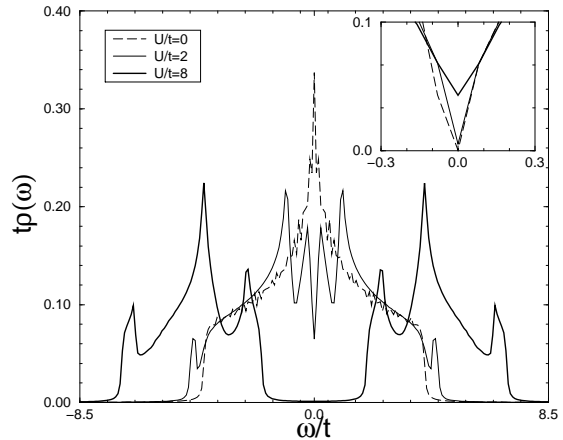,height=6.5cm,angle=-90}}
\caption{
DOS for the $2D$ Falicov-Kimball model for
several values of $U/t$ at half-filling, with our method.
It is always positive definite.
The inset shows the behavior of the $ U/t=2$ gap at $\omega=0$ when changing
the artificial broadening $\eta$ of the $\delta$ functions.
$\eta/t= 5.10^{-2}$ thick line, $\eta/t=5.10^{-3}$ thin line and
$\eta/t=2.10^{-4}$ dashed line.
}
\end{figure}                        
Fig.2 shows the density of states (DOS) $\rho(\omega)$ 
obtained by our method for several values of $U/t$.  Firstly,  
notice that this function is always positive definite, in contrast
to results of ref. \cite{SI}. This is a positive feature of the present
method.  We should stress that we are interested in the effect of coupling of
self-energy to $1/D$ fluctuations, which, for the FKM, are completely
static.   
 The effect of
the interaction is very clear. Starting from the $2D$ free fermions
value at $U/t=0$, it opens a gap at $\omega=0$. 
The critical $(U/t)_{c}$ for opening up a gap is very small. As a matter
 of fact, working with a $L\times L$ lattice 
we cannot resolve energy details smaller than $2\pi v_F/L \simeq 0.2$ here.
However, playing with the parameters of the problem, we see a real 
gap with $U/t$ as low as $0.1$.  This result agrees with the exact result 
on the $2D$ FKM at weak coupling\cite{lember}, where the gap and the related 
checkerboard 
order arises from the nested FS at $n=1$. The inset of fig.2
shows the behavior of the DOS at low frequency for $U/t=2$, obtained 
by changing 
the broadening of the delta functions to get a better precision. 
The features observed at higher frequencies are 
related to the unperturbed $2D$ band structure and are absent in 
infinite $D$. 

The computed $\chi_{\downarrow}({\bf q})$, which contains information 
about long-range order (LRO), shows interesting behavior; along the 
zone-diagonal, at the point  
(${\bf k} =[\pi,\pi]$), it is a constant, scaling approximately 
with lattice size.  This is the expected behavior in the broken-symmetry 
phase \cite{farkasovsky}, and implies that the $n=1$ ground state is 
characterized by checkerboard LRO of the down-spin electrons 
\cite{freericks}. In ref.\cite{farkasovsky}, the behavior of 
$\chi_\downarrow({\bf q})$ can be studied only in $1D$, because of finite 
size restrictions. In contrast, our approach yields unambiguous conclusions 
in the thermodynamic limit in $2D$.  Again, this is in clear agreement 
with the exact result\cite{lember}.
\begin{figure}
\centerline{\psfig{file=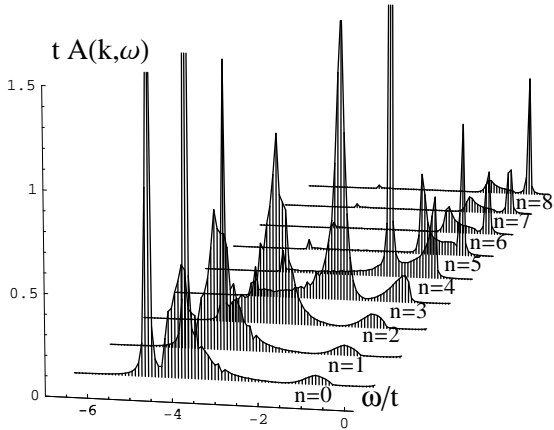,height=6.5cm,angle=0}}
\caption{The spectral function $A_\uparrow({\bf k},\omega)$ for
the $2D$ Falicov-Kimball model at half-filling, for $U/t=4$, and {\bf k}
along the zone diagonal, ${\bf k}=(n\pi/8,n\pi/8)$, $n$ going from 0
to 8.
}
\end{figure}   
The greatest advantage of our method, however, is that it allows a 
detailed study of ${\bf k}$-dependent (at the order of $1/D$) effects 
in the one particle spectral function 
$A({\bf k},\omega)=-{\rm Im}{G({\bf k},\omega)}/\pi$.  
In fig.3, we show $A({\bf k},\omega)$ for ${\bf k}$ along 
the zone-diagonal from ${\bf k}={\bf 0}$ to ${\bf k}=(\pi,\pi)$ for a 
representative value of $U=4t$. We notice that $A({\bf k},\omega)$ 
satisfies all the symmetry properties consistent with those expected 
from pure nearest neighbor hopping on bipartite lattices, and with 
particle-hole symmetry: $A({\bf k}_0,\omega)=A({\bf k}_0,-\omega)$ for 
${\bf k}_0=[\pi/2,\pi/2]$, while along the zone diagonal 
$A({\bf k}_0+{\bf \delta},\omega)=A({\bf k}_0-{\bf \delta},-\omega)$, 
$\forall {\bf \delta}$ along the zone diagonal.  We show only the
$A({\bf k},\omega)$ for $\omega < 0$; the $\omega > 0$ part can be obtained 
from the symmetry property mentioned above.  We have checked that our 
calculations are fully consistent with the above property. 
 Moreover, we see 
that since 
${\bf Q}=[\pi,\pi]$ is a nesting vector in $2D$, the above property of 
$A({\bf k},\omega)$ with ${\bf \delta}=[\pi/2,\pi/2]$ leads to a natural 
explanation of the ``shadow-band'' features observable in ARPES.  
It would be very interesting to see whether they survive away from $n=1$.  

It is instructive to compare our results to those obtained by Velick\'y {\it 
et. al}\cite{VKE} in their pioneering paper on CPA.  
CPA is equivalent to the exact $D=\infty$ result for the FKM, 
and so a comparison with \cite{VKE} allows us to study the effects of 
$1/D$ effects on $A({\bf k},\omega)$. A comparison of our data with 
results for identical parameters from \cite{VKE} shows that the coupling 
to $1/D$ fluctuations induces new features in $A({\bf k},\omega) $ 
compared to those observed in DMFT.
In $D=\infty$, $A({\bf k},\omega)$ has a two-peak structure with the 
dispersion controlled solely by the free bandstructure. In our approach,
this is modified because of the extra ${\bf k}$-dependence coming 
through the $B_{{\bf k}\downarrow} $. 
This shows that a formalism capable of explicitly 
treating intersite correlations permits one to access ${\bf k}$-dependent 
features in $A({\bf k},\omega)$, so we suggest that this method can 
be fruitfully applied to compute ARPES lineshapes in correlated systems.
A more detailed application to spectroscopy would require the use of the actual
bandstructure DOS, and is left for future work.

In conclusion, we have developed a simple, physically appealing way to study
the effects of non-local spatial fluctuations on the single particle spectral 
properties. We have applied the formalism to compute the single particle 
spectral function and the DOS for the $2D$ FKM and have obtained
 results in good 
accordance with what is known from the exact solution.  
Extensions of the work to look
at the metallic phase in the FKM off $n=1$, as well as for the Hubbard model, 
are being studied and will be reported separately.

Mukul S.Laad wishes to thank
 the Alexander von Humboldt Stiftung for financial support during the time this project was started.  We
 thank Prof. P.Fulde for advice and hospitality at the MPI, Dresden.

%\vspace{5mm}

\noindent $^1$ e-mail: mukul@thp.uni-koeln.de

\noindent $^2$ e-mail: vdb@irsamc2.ups-tlse.fr

\end{multicols}

\end{document}